\documentclass{article}

\usepackage{lineno,hyperref}
\modulolinenumbers[5]
\usepackage{amsmath,amsthm,amssymb}
\usepackage{mathabx}
\usepackage{url}
\usepackage{ucs}
\usepackage{braket}
\usepackage{natbib}
\usepackage{authblk}
\usepackage{graphicx}
\usepackage{abstract}

\newcommand{\ignore}[1]{}
\newcommand{\given}{\,|\,}
\newcommand{\params}{\theta}
\newcommand{\as}{\star}

\title{Using the Expectation Maximization Algorithm with Heterogeneous Mixture Components for the Analysis of Spectrometry Data}

\author[1]{Dominik Kopczynski}
\author[2]{Sven Rahmann}

\affil[1]{Collaborative Research Center SFB~876, TU~Dortmund, Germany}
\affil[2]{Genome Informatics, Institute of Human Genetics, Faculty of Medicine, University of Duisburg-Essen, and University Hospital Essen, Germany}

\begin{document}
\maketitle
\begin{abstract}
Coupling a multi-capillary column (MCC) with an ion mobility (IM) spectrometer (IMS) opened a multitude of new application areas for gas analysis, especially in a medical context, as volatile organic compounds (VOCs) in exhaled breath can hint at a person's state of health.
To obtain a potential diagnosis from a raw MCC/IMS measurement, several computational steps are necessary, which so far have required manual interaction, e.g., human evaluation of discovered peaks.
We have recently proposed an automated pipeline for this task that does not require human intervention during the analysis.
Nevertheless, there is a need for improved methods for each computational step.
In comparison to gas chromatography / mass spectrometry (GC/MS) data, MCC/IMS data is easier and less expensive to obtain, but peaks are more diffuse and there is a higher noise level.
MCC/IMS measurements can be described as samples of mixture models (i.e., of convex combinations) of two-dimensional probability distributions.
So we use the expectation-maximization (EM) algorithm to deconvolute mixtures in order to develop methods that improve data processing in three computational steps: denoising, baseline correction and peak clustering.
A common theme of these methods is that mixture components within one model are not homogeneous (e.g., all Gaussian), but of different types.
Evaluation shows that the novel methods outperform the existing ones.
We provide Python software implementing all three methods and make our evaluation data available at \url{http://www.rahmannlab.de/research/ims}.
\end{abstract}



\section{Introduction}
\label{sec:intro}

\paragraph*{Technology background}
An ion mobility (IM) spectrometer (IMS) measures the concentration of volatile organic compounds (VOCs) in the air or exhaled breath by ionizing the compounds, applying an electric field and measuring how many ions drift through the field after different amounts of time.
A multi-capillary column (MCC) can be coupled with an IMS to pre-separate a complex sample by retaining different compounds for different times in the columns (according to surface interactions between the compound and the column).
As a consequence, compounds with the same ion mobility can be distinguished by their distinct retention times.

Recently, the MCC/IMS technology has gained importance in medicine, especially in breath gas analysis, as VOCs may hint at certain diseases like lung cancer, chronic obstructive pulmonary disease (COPD) or sarcoidosis~\citep{Westhoff/etal/2009a, bodeker2008peakcomp, bunkowski/etal/2009a, westhoff2010differentiation}.

\begin{figure*}[t]\centering
\includegraphics[width=0.98\linewidth]{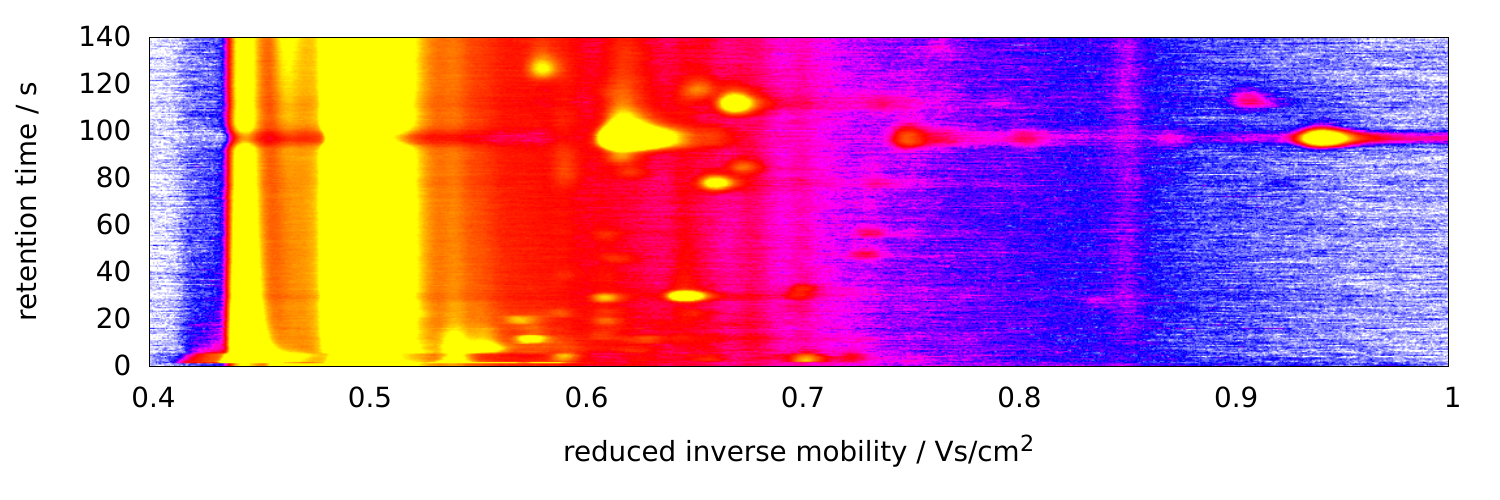}
\caption{
Visualization of a raw IMSC as a heat map.
X-axis: reduced inverse mobility $1/K_0$ in $\text{Vs}/\text{cm}^2$; 
y-axis: retention time~$r$ in seconds; 
signal: white (lowest) $<$ blue $<$ purple $<$ red $<$ yellow (highest), reactant ion peak (RIP) at $0.48\,\text{Vs}/\text{cm}^2$.
}
\label{fig:visualized-imsc}
\end{figure*}

A typical MCC/IMS measurement takes about ten minutes.
Within this time the MCC pre-separates the sample.
An IM spectrum is captured periodically every 100~ms.
The aligned set of captured IMS spectra is referred to an IM spectrum-chromatogram (IMSC) which consists of an~$|R| \times |T|$ matrix~$S = (S_{r,t})$, where~$R$ is the set of retention time points (whenever an IM spectrum is captured, measured in seconds), and~$T$ is the set of drift time points (whenever an ion measurement is made; in milliseconds).
To remove the influences of pressure, ambient temperature or drift tube size, a normalized quantity is used instead of drift time, namely the reduced mobility~$K_0$ with units of~$\text{cm}^2\text{V}^{-1}\text{s}^{-1}$, as described by~\cite{eiceman/2010a}.
(Reduced) mobility is inversely proportional to drift time, so we consider the reduced inversed mobility (RIM)~$1/K_0$ with units of~$\text{Vs}/\text{cm}^2$.
RIM and drift time are proportional, with the proportionality constant depending on the above external quantities.
As not mentioned otherwise, in the following we use $U=4830\,\text{V}$ and $\ell=12\,\text{cm}$ which corresponds to the voltage and length of our IMS drift tube.
We assume that all time points (or RIMs) are equidistant; 
so we may work with matrix indices~$r\in\{1,\dots,|R|\}$ and~$t\in\{1,\dots,|T|\}$ for convenience.
On average, an IM spectrum takes about $50\,\text{ms}$, corresponding to $1.45\,\text{Vs}/\text{cm}^2$ and an IMSC about~$600\,\text{s}$.
The signal values of an IMSC are digitized by an analog-digital converter with a precision of~12~bits.
Since the device can operate in positive and negative mode, the values range between~$-2048$ and~$2047$.

Figure~\ref{fig:visualized-imsc} visualizes an IMSC as a heat map.
Areas with a high signal value are referred to as \emph{peaks}.
A peak is caused by the presence (and concentration) of a certain compound; the peak position~$(r,t)$ indicates which compound is present, and the peak volume contains information about the concentration.

An inherent feature of IMS technology is that the drift gas is ionized, too, which results in a ``peak'' that is present at each retention time at a RIM of~$0.48~\text{Vs}/\text{cm}^2$ (Figure~\ref{fig:visualized-imsc}). 
It is referred to as the reactant ion peak (RIP).

\paragraph*{Related work and novel contributions}
A typical work flow from a raw IMSC to a ``diagnosis'' or classification of the measurement into one of two (or several) separate classes generally proceeds along the following steps described by \cite{daddario/etal/2014a}:
pre-processing, peak candidate detection, peak picking, and parametric peak modeling.

All methods of this paper are adaptations of the expectation-maximization (EM) algorithm, modified for their particular task.
The EM~algorithm (introduced by \cite{Dempster/etal/1977a}) is a statistical method frequently used for deconvolving distinct components in mixture models and estimating their parameters.
We summarize its key properties in Section~\ref{sec:algo:em}.

We previously used the EM~algorithm for peak modeling, at the same time decomposing a measurement into separate peaks. 
We introduced a model that describes the shape of a peak with only seven parameters using a two-dimensional shifted inverse Gaussian distribution function with an additional volume parameter \citep{Kopczynski/etal/2012a}.

We also evaluated different peak candidate detection and peak picking methods, comparing for example manual picking by an expert with state of the art tools like IPHEx \citep{bunkowski_2011}, VisualNow \citep{bader2007reduction}, and our own methods \citep{hauschild2013eval}.

In this work, we focus on pre-processing.
Pre-processing is a crucial step because it determines the difficulty and the accuracy with which peak candidates (and peaks) can be identified and correctly modeled.
It consists of several sub-tasks: denosing, baseline correction and smoothing.
We discuss novel methods for denoising with integrated smoothing (Section~\ref{sec:denoising}) and for baseline correction (Section~\ref{sec:baseline}).

A second focus of this work is on finding peaks that correspond to each other (and hence to the same measured compound) in several measurements of a dataset.
We introduce an EM-based clustering method (Section~\ref{sec:clustering}).
An accurate clustering is important for determining feature vectors for classification.
As the detected location of a peak may differ between several measurements, a clustering approach across measurements suggests itself.
Several clustering algorithms like~$K$-means (first introduced by~\cite{macqueen1967kmeans}) or hierarchical clustering have the disadvantage that they need a fixed number of models or a threshold for the density within a cluster.
In practice, these algorithms are executed several times with an increasing number of clusters and take the best result with respect to a cost function penalized with model complexity.
DBSCAN~\citep{ester/etal/1996} is a clustering method which does not require a fixed cluster number.

We demonstrate that our proposed EM variants outperform existing methods for their respective tasks in Section~\ref{sec:eval} and conclude the paper with a brief discussion.
\section{Algorithms}
\label{sec:algo}

This section describes our adaptations of the EM~algorithm (summarized in Section~\ref{sec:algo:em}) for denoising (Section~\ref{sec:denoising}), baseline correction (Section~\ref{sec:baseline}) and peak clustering across different measurements (Section~\ref{sec:clustering}).
The first two methods use heterogeneous model components, while the last one dynamically adjusts the number of clusters.
For each algorithm, we present background knowledge, the specific mixture model, the choice of initial parameter values, the maximum likelihood estimators of the M-step (the E-step is described in Section~\ref{sec:algo:em}), and the convergence criteria.
For peak clustering, we additionally describe the dynamic adjustment of the number of components.
The algorithms are evaluated in Section~\ref{sec:eval}.

\subsection{The EM Algorithm for Mixture Models with Heterogeneous Components}
\label{sec:algo:em}

In all subsequent sections, variations of the EM~algorithm \citep{Dempster/etal/1977a} for mixture model deconvolution are used.
Here we summarize the algorithm and describe the E-step common to all variants.

A fundamental idea of the EM~algorithm is that the observed data~$x$ is viewed as a sample of a mixture (convex combination)~$f$ of probability distributions,
\[ f(x \given \params) = \sum_{c=1}^C\, \omega_c\, f_c(x \given \params_c), \]
where $c$~indexes the~$C$~different component distributions~$f_c$, where $\params_c$ denotes all parameters of distribution~$f_c$, and $\params=(\params_1, \dots, \params_c)$ is the collection of all parameters.
The mixture coefficients $\omega_c$ satisfy $\omega_c\geq 0$ for all~$c$, and $\sum_{c}\, \omega_c = 1$.

We point out that, unlike in most applications, in our case the probability distributions $f_c$ are of different types, e.g., a uniform and a Gaussian one.

The goal of mixture model analysis is to estimate the mixture coefficients~$\omega = (\omega_c)$ and the individual model parameters~$\params = (\params_c)$, whose number and interpretation depends on the parametric distribution~$f_c$.

Since the resulting maximum likelihood parameter estimation problem is non-convex, iterative locally optimizing methods such as the Expectation Maximization (EM) algorithm are frequently used.
The EM~algorithm consists of two repeated steps:
The E-step (expectation) estimates the expected membership of each data point in each component and then the component weights~$\omega$, given the current model parameters~$\params$.
The M-step (maximization) estimates maximum likelihood parameters~$\params_c$ for each parametric component~$f_c$ individually, using the expected memberships as hidden variables that decouple the model.
As the EM~algorithm converges towards a local optimum of the likelihood function, it is important to choose reasonable starting parameters for~$\theta$.

\paragraph{E-step}
The E-step is independent of the specific component distribution types and always proceeds in the same way, so we summarize it here once, and focus on the specific M-step in each of the following subsections.
To estimate the expected membership~$W_{i,c}$ of data point~$x_i$ in each component~$c$, the component's relative probability at that data point is computed, i.e.,
\begin{equation}
  W_{i,c} = \frac{\omega_c\, f_c(x_i \given \params_c)}{\sum_{k}\, \omega_k\, f_k(x_i \given \params_k)},
\label{eq:Estep}
\end{equation}
such that~$\sum_c\, W_{i,c} = 1$ for all~$i$.
Then the new component weight estimates~$\omega^{\as}_c$ are the averages of~$W_{i,c}$ across all data points,
\begin{equation}
  \omega^{\as}_c = \frac{1}{n} \sum_{i=1}^n\, W_{i,c},
\label{eq:weights}  
\end{equation}
where~$n$ is the number of data points.

\paragraph{Convergence}
After each M-step of an EM~cycle, we compare $\theta_{c,q}$ (old parameter value) and $\theta^*_{c,q}$ (updated parameter value), where~$q$ indexes the elements of~$\theta_c$, the parameters of component~$c$.
We say that the algorithm has converged when the relative change
\[
\kappa  :=
    \frac{|\theta_{c,q}^*  - \theta_{c,q}|}{\max \left( |\theta_{c,q}^*|,|\theta_{c,q}| \right)}
\]
drops below the threshold~$\varepsilon:=0.001$, corresponding to~$0.1\%$ precision, for all~$c,q$.
(If $\theta_{c,q}^* = \theta_{c,q} = 0$, we set $\kappa := 0$.)

\subsection{Denoising}
\label{sec:denoising}

\paragraph*{Background}
A major challenge during peak detection in an IMSC is to find peaks that only slightly exceed the background noise level.

As a simple method, one could declare each point~$(r,t)$ as a peak whose intensity~$S_{r,t}$ exceeds a given threshold.
In IMSCs, peaks become wider with increasing retention time, while their volume remains constant, so their height shrinks, while the intensity of the background noise remains at the same level.
So it is not appropriate to choose a constant noise level threshold, as peaks at high retention times may be easily missed.

To determine whether the intensity~$S_{r,t}$ at coordinates~$(r,t)$ belongs to a peak region or can be solely explained by background noise, we propose a method based on the EM~algorithm.
It runs in~$\mathcal{O}(\tau |R||T|)$ time where~$\tau$ is the number of EM~iterations.
Before we explain the details of the algorithm, we mention that it does not run on the IMSC directly, but on a smoothed matrix~$A$ containing local averages from a window with margin~$\rho$;
\begin{equation*}
	A_{r, t} \coloneq \frac{1}{(2\rho+1)^2} \cdot \sum\limits_{r' = r - \rho}^{r + \rho} \sum\limits_{t' = t - \rho}^{t + \rho} S_{r', t'}
\end{equation*}
for all~$r\in\{1,\dots,|R|\}$, $t\in\{1,\dots,|T|\}$.
Since the borders of an IMSC do not contain important information, we deal with boundary effects by computing~$A_{r,t}$ in those cases as averages of only the existing matrix entries.

To choose the smoothing radius~$\rho$, we consider the following argument.
For distinguishing two peaks, \cite{bodeker2008peak} introduced a minimum distance in reduced inverse mobility of~$\Delta t \coloneq 0.003\,\text{Vs}/\text{cm}^2$.
In our datasets (2500 drift times with a maximal value of reduced inverse mobility of~$1.45\,\text{Vs}/\text{cm}^2$), this corresponds to~$5.17 \approx 5$ index units, so we use~$\rho=4$ index units to avoid taking to much noise into consideration.

\paragraph{Mixture model}
Based on observations of IMSC signal intensities, we assume that
\begin{itemize}
\item the noise intensity has a Gaussian distribution over low intensity values with mean~$\mu_{\text{N}}$ and standard deviation~$\sigma_{\text{N}}$,
\[ f_{\text{N}}(s \given \mu_{\text{N}}, \sigma_{\text{N}}) 
   = \frac{1}{\sqrt{2 \pi} \, \sigma_{\text{N}}} \cdot \exp \big( -(s - \mu_{\text{N}})^2 / (2\, \sigma_{\text{N}}^2) \big)
\]
\item the true signal intensity has an Inverse Gaussian distribution with mean $\mu_{\text{S}}$ and shape parameter~$\lambda_{\text{S}}$, i.e.,
\[ f_{\text{S}}(s \given \mu_{\text{S}}, \lambda_{\text{S}}) 
   = \sqrt{\lambda_{\text{S}} / (2 \pi s^3)} \cdot \exp \big( -\lambda_{\text{S}} (s - \mu_{\text{S}})^2 / (2 \mu_{\text{S}}^2 s) \big)
\]
\item there is an unspecific background component which is not well captured by either of the two previous distributions; we model it by the uniform distribution over all intensities,
\[ f_\text{B}(s) = (\max(S) - \min(S))^{-1},
\]
and we expect the weight~$\omega_\text{B}$ of this component to be close to zero in standard IMSCs, a deviation indicating some anomaly in the measurement.
\end{itemize}

\begin{figure}[t]
  \centering
  \includegraphics[width=0.98\textwidth]{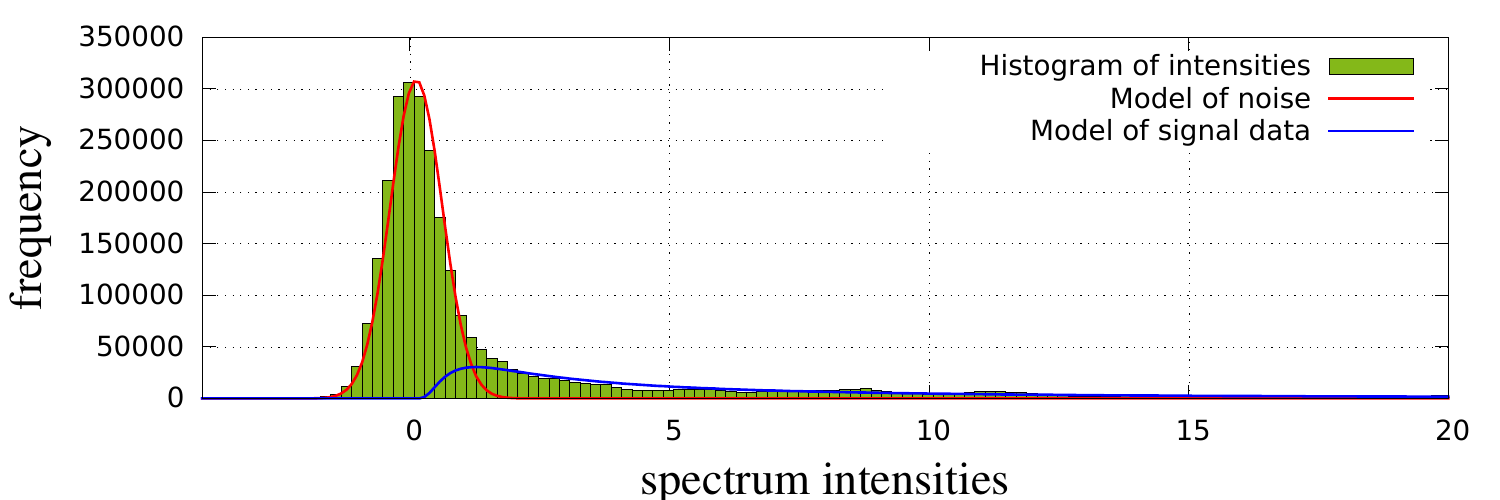}
  \caption{Histogram of a smoothed IMSC~$A$ (green bars) and estimated distribution of the noise component (red line) and of the signal component (blue line). 
Parameters for both components were estimated with the EM~algorithm.}
  \label{fig:denoisingmodels}
\end{figure}

We interpret the smoothed observed IMSC~$A$ as a sample of a mixture of these three components with unknown mixture coefficients.
To illustrate this approach, consider Figure~\ref{fig:denoisingmodels}, which shows the empirical intensity distribution of an IMSC (histogram), together with the estimated components (except the uniform distribution, which has the expected coefficient of almost zero).

It follows that there are six independent parameters to estimate: $\mu_{\text{N}}$, $\sigma_{\text{N}}$, $\mu_{\text{S}}$, $\lambda_{\text{S}}$ and weights $\omega_{\text{N}}, \omega_{\text{S}},\omega_{\text{B}}$ (noise, signal, background, where $\omega_{\text{B}} = 1 - \omega_{\text{N}} - \omega_{\text{S}}$).

\paragraph{Initial parameter values}
As the first and last~$10\%$ of data points in each spectrum can be assumed to contain no signal, we use their intensities' empirical mean and standard deviation as starting values for~$\mu_{\text{N}}$ and~$\sigma_{\text{N}}$, respectively.
The initial weight of the noise component is set to cover most points covered by this Gaussian distribution, i.e., $\omega_{\text{N}} := |\{(r,t) \in R\times T \given A_{r,t} \leq \mu_{\text{N}} + 3\, \sigma_{\text{N}}\}| \; /\; (|R||T|)$.

We assume that almost all of the remaining weight belongs to the signal component, thus $\omega_{\text{S}} = (1 - \omega_{\text{N}}) \cdot 0.999$, and $\omega_{\text{B}} = (1 - \omega_{\text{N}}) \cdot 0.001$. 

To obtain initial parameters for the signal model, let $I' := \{(r,t) \in R\times T \given A_{r,t} > \mu_{\text{N}} + 3 \, \sigma_{\text{N}}\}$ (the complement of the intensities that are initially assigned to the noise component).
We set $\mu_{\text{S}} = \big( \sum_{(r,t) \in I'}\, A_{r,t} \big) / |I'|$ and $\lambda_{\text{S}} = (\sum_{(r,t) \in I'}\, (1/A_{r,t} - 1 / \mu_{\text{S}}))^{-1}$ (which are the maximum likelihood estimators for Inverse Gaussian parameters).

\paragraph*{Maximum likelihood estimators}
In the maximization step (M-step) we estimate maximum likelihood parameters for the non-uniform components.
In all sums,~$i=(r,t)$ extends over the whole matrix index set~$R\times T$.
\begin{align}
\mu_{c} &= \frac{\sum_{i}\, W_{i, c} \cdot A_i}{\sum_{i}\, W_{i, c}},  \qquad c\in\{\text{N},\text{S}\},
\label{equ:mstepmueDN} \\
\sigma^2_{\text{N}} &= \frac{ \sum_{i}\, W_{i, \text{N}} \cdot (A_i - \mu_{\text{N}})^2 }{ \sum_{i}\, W_{i, \text{N}} }, \label{equ:mstepsigmaDN} \\
\lambda_{\text{S}} &= \frac{ \sum_{i}\, W_{i, \text{S}} }{ \sum_{i}\, W_{i, \text{S}} \cdot (1 / A_i - 1 / \mu_{\text{S}}) }. \label{equ:msteplambdaDN}
\end{align}
\paragraph*{Final step}
After convergence (8--10 EM~loops in practice), the denoised signal matrix~$S^+$ is computed as follows:
\[ S^+_{i} \coloneq S_{i} \cdot (1 - W_{i,\text{N}}) \text{ for all } i\in R\times T.
\]
\subsection{Baseline Correction}
\label{sec:baseline}

\paragraph*{Background}
In an IMSC, the RIP with its long tail interferes with peak detection; it is present is each spectrum and hence called the \emph{baseline}.
The goal of this section is to remove the baseline and better characterize the remaining peaks.

\begin{figure}[t]\centering
\includegraphics[width=0.48\columnwidth]{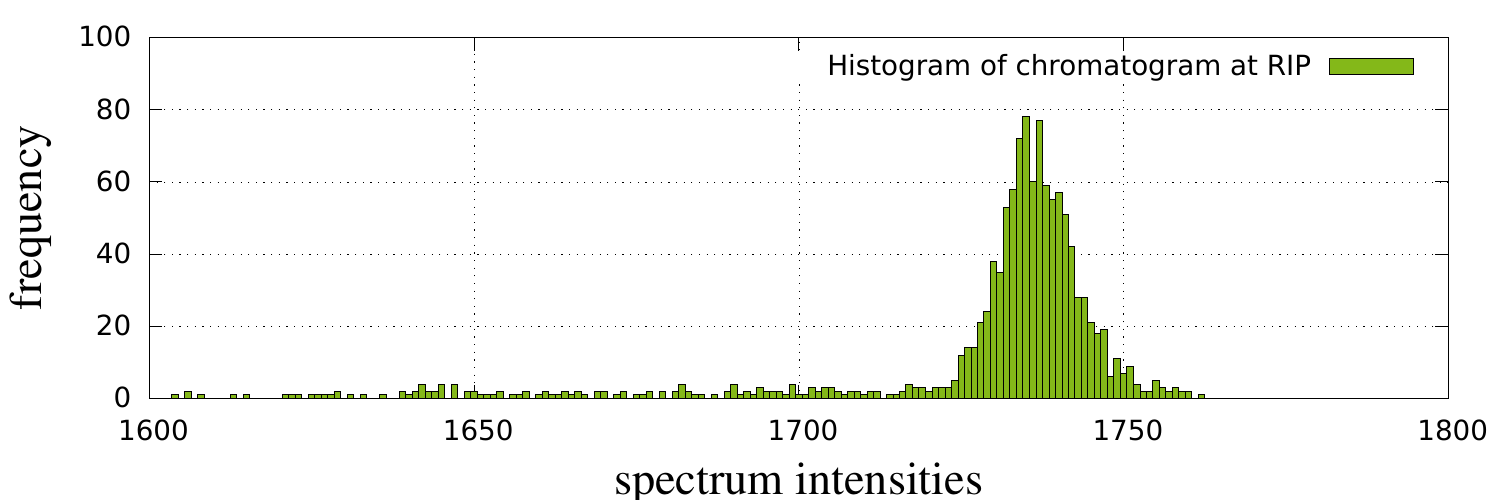}
\includegraphics[width=0.48\columnwidth]{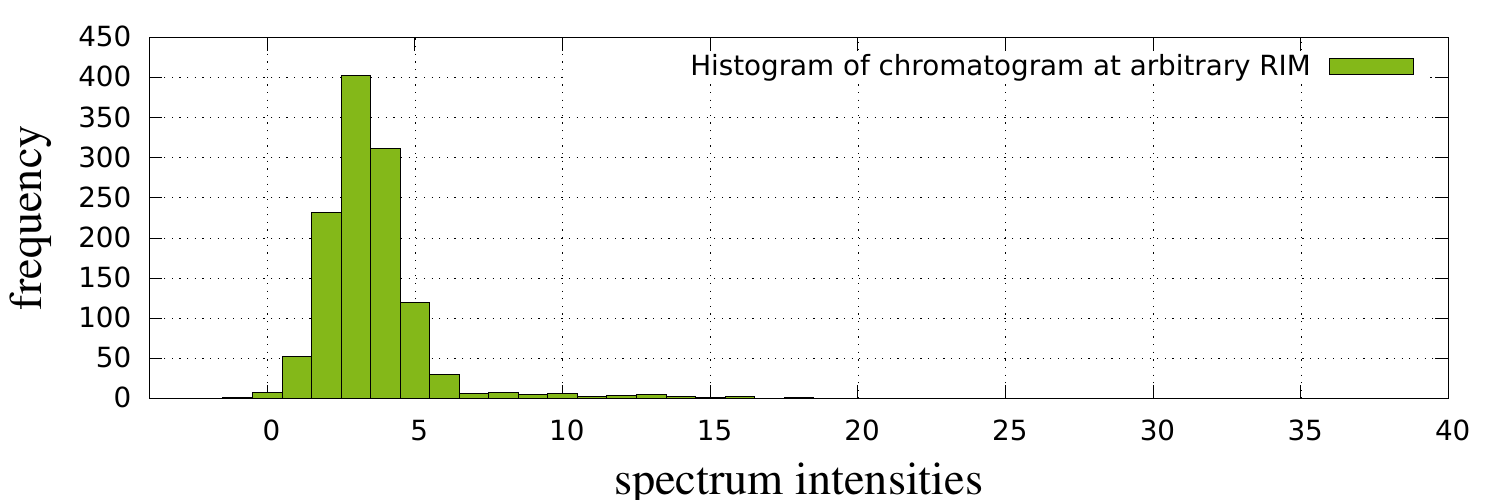}
\caption{Two typical histograms of chromatogram signal intensities.\
Top: Most data points of a RIP chromatogram consist of high values with high variance.
Bottom: Far away from the RIP, only few values right of background noise are produced by peaks.
}
\label{fig:histogram-chromatograms}
\end{figure}

We consider every chromatogram (column of the matrix shown in Figure~\ref{fig:visualized-imsc}) separately.
The idea is to consider intensities that appear at many retention times as part of the baseline.
By analyzing the histogram~$H_t$ of chromatogram~$S_{\cdot, t}$ (with bin size~$1$, since signal values are integers), we observe that frequently occurring signals that are produced by the IM device itself or by drift gas, build the highest peak in the histogram, consider Figure~\ref{fig:histogram-chromatograms} (top).
On the other hand, histograms of chromatograms that are only negligibly influenced by the RIP have a peak in the range of the background noise mean, see Figure~\ref{fig:histogram-chromatograms} (bottom).

\paragraph{Mixture model}
We make the following assumption based on observations of chromatogram intensities:
\begin{itemize}
\item The intensities belonging to the baseline are normally distributed around their mean,
\[ f_\text{B}(s \given \mu, \sigma)
   = \frac{1}{\sqrt{2 \pi}\, \sigma} \cdot \exp \big(-(s - \mu)^2 / (2\,\sigma^2) \big),
\]
\item The remaining intensities belong to the signal of interest and can have any value above the baseline, so they are modeled by a uniform distribution between the minimum value~$m:=\min_r\, S_{r,t}$ and maximum value~$M:=\max_r\, S_{r,t}$ in the chromatogram~$S_{\cdot,t}$ at drift time~$t$,
\[ f_\text{S}(s) = \frac{1}{M-m}.
\]
\end{itemize}

\paragraph*{Initial parameter values}
The start parameter for~$\mu$ is the most frequent intensity in the chromatogram (the mode of the histogram); we also set~$\sigma=1$ and $\omega_\text{B}=0.9$, $\omega_\text{S}=1-\omega_\text{B}$.

\paragraph*{Maximum likelihood estimators}
The new values for mean and standard deviation of~$f_\text{B}$ are estimated by the standard maximum likelihood estimators, weighted by component membership.
The following formulas apply to a single chromatogram~$S_{\cdot,t}$.
\begin{align}
	\mu &= \frac{ \sum_{i\in R} W_{i,\text{B}} \cdot S_{i,t}}{ \sum_{i\in R} W_{i, \text{B}} } 
	\label{equ:mstepmueRC}, \\
	\sigma^2 &= \frac{\sum_{i \in R} W_{i,\text{B}} \cdot (\mu - S_{i, t})^2 }{\sum_{i \in R} W_{i,\text{B}}}. 
	\label{equ:mstepsigmaRC}
\end{align}

\paragraph*{Final step}
When the parameters converge, the baseline intensity for~$S_{\cdot,t}$ is estimated at~$B_t := \mu + 2\sigma$ (note that we omitted the index~$t$ for~$\mu$ and~$\sigma$, as each chromatogram is processed independently).
This baseline value is subtracted from each intensity in the chromatogram, setting resulting negative intensities to zero.
In other words, the new matrix is $S^+_{r,t} := \max\{ S_{r,t} - B_t, 0\}$.

\subsection{Clustering}
\label{sec:clustering}

\paragraph*{Background}
Peaks in several measurements are described by their location in retention time and reduced inverse mobility (RIM).
Let~$X$ be a union set of peak locations from different measurements with~$|X| = n$ being the number of peaks and~$X_{i,\text{R}}$ the retention time of peak~$i$ and~$X_{i,\text{T}}$ its RIM.
We assume that due to the slightly inaccurate capturing process, a peak (produced by the same compound) that appears in different measurements has slightly shifted retention time and RIM.
We introduce a clustering approach using standard 2-dimensional Gaussian mixtures, but with dynamically adjusting the number of clusters in the process.

\paragraph{Mixture model}
We assume that the measured retention times and RIMs belonging to peaks from the same compound are independently normally distributed in both dimensions around the (unknown) component retention time and RIM.
Let $\theta_j := (\mu_{j,\text{R}}, \sigma_{j,\text{R}}, \mu_{j,\text{T}}, \sigma_{j,\text{T}})$ be the parameters for component~$j$, and let $f_j$ be a two-dimensional Gaussian distribution for a peak location $x=(x_\text{R}, x_\text{T})$ with these parameters,
\[ f_j(x \given \theta_j) 
   = \frac{1}{2 \pi \, \sigma_{j,\text{R}} \, \sigma_{j,\text{T}}}
   \, \exp \left( 
   -\frac{(x_{\text{T}} - \mu_{j,\text{T}})^2}{2\,\sigma_{j,\text{T}}^2}
   -\frac{(x_{\text{R}} - \mu_{j,\text{R}})^2}{2\,\sigma_{j,\text{R}}^2}
   \right).
\]
The mixture distribution is $f(x) = \sum_{j=1}^C\, \omega_j\, f_j(x\given \theta_j)$ with a yet undetermined number~$C$ of clusters.
Note that there is no ``background'' model component.

\paragraph*{Initial parameter values}
In the beginning, we initialize the algorithm with as many clusters as peaks, i.e., we set~$C:=n$.
This assignment makes a background model obsolete, because all peaks are assigned to at least one cluster.
All clusters get as start parameters for $\mu_{j,\text{R}}, \mu_{j,\text{T}}$ the original retention time and  RIM of peak location~$X_j$, respectively, for~$j = 1,\dots,n$.

Remark that we are using in this description not the indices but the actual measures.
We set $\sigma_{j,\text{T}} := 0.003\,\text{Vs} / \text{cm}^2$ and $\sigma_{j,\text{R}} := (0.1 \, X_{j,\text{R}} + 3\,\text{s}) / 3$ according to the peak characterizations by \cite{bodeker2008peak}, dividing by~3 to let~$3\,\sigma$ since due to the strong skewed peaks in retention time the area under the curve is asymmetric.

\paragraph*{Dynamic adjustment of the number of clusters}
After computing weights in the E-step, but before starting the M-step, we dynamically adjust the number of clusters by merging clusters whose centers are close.
Every pair~$j < k$ of clusters is compared in a nested for-loop.
When $|\mu_{j,\text{T}} - \mu_{k,\text{T}}| < 0.003 \text{ Vs} / \text{cm}^2$ and $|\mu_{j,\text{R}} - \mu_{k,\text{R}}| < 0.001 \cdot \max\{\mu_{j, \text{R}},\mu_{k, \text{R}}\} + 3\,\text{s}$, then clusters~$j$ and~$k$ are merged by summing the weights $\omega^+:=\omega_j+\omega_k$ and $W_{i,+}:=W_{i,j}+W_{i,k}$ for all~$i$, and these are assigned to the location of the cluster with larger weight.
(The re-computation of the parameters happens immediately after merging in the maximization step.)
The comparison order may matter in rare cases for deciding which peaks are merged first, but since new means and variances are computed, possible merges that were omitted in the current iteration, will be performed in the next iteration.
This merging step is applied first time in the second iteration, since the cluster means need at least one iteration to move towards each other.

\paragraph*{Maximum likelihood estimators}
The maximum likelihood estimators for mean and variance of a two-dimensional Gaussian are the standard ones, taking into account the membership weights,
\begin{align}
\mu_{j,d}
  &= \frac{\sum_{i=1}^n\, W_{i, j} \cdot X_{i, d}}{\sum_{i=1}^n\, W_{i, j}}, 
  & d\in\{\text{T,R}\}, \\
\sigma_{j,d}^2
  &= \frac{\sum_{i=1}^n\, W_{i, j} \cdot (X_{i, d} - \mu_{j, d})^2} {\sum_{i=1}^n\, W_{i, j}}, 
  & d\in\{\text{T,R}\}, 
\end{align}
for all components~$j=1,\dots,C$.

One problem using this approach emerges from the fact that initially each cluster contains only one peak, leading to an estimated variance of zero in many cases.
To prevent this, minimum values are enforced such that  $\sigma_{j,\text{T}} \geq 0.003\,\text{Vs}/\text{cm}^2$ and $\sigma_{j,\text{R}} \geq (0.1\, \mu_{j,\text{R}} + 3\,\text{s}) / 3$ for all~$j$.

\paragraph*{Final step}
The EM~loop terminates when no merging occurs and the convergence criteria for all parameters are fulfilled.
The resulting membership weights determine the number of clusters as well as the membership coefficient of peak location~$X_i$ to cluster~$j$.
If a hard clustering is desired, the merging step has to be protocoled.
At the beginning all peak indexes are singletons within their own sets.
By merging, the sets of both peaks are merged.

%

\section{Evaluation}
\label{sec:eval}

In this section, we evaluate our algorithms against existing state-of-the-art ones on simulated data.
We first discuss general aspects of generating simulated IMSCs (Section~\ref{sec:eval:general}) and then report on the evaluation results for denoising (Section~\ref{sec:eval:denoising}), baseline correction (Section~\ref{sec:eval:baseline}) and peak clustering (Section~\ref{sec:eval:clustering}).

\subsection{Data Generation and Similarity Measure}
\label{sec:eval:general}

Since we do not have ``clean'' real data, we decided to simulate IMSCs and add noise with the same properties as observed in real IMS datasets.
We generate simulated IMSCs of~$1200$ retention time points and~$2500$ RIM points with several peaks (see below), subsequently add noise (see below), apply our and competing algorithms and compare the resulting IMSCs with the original simulated one.

\paragraph{Simulating IMSCs with peaks}
A peak in an IMSC can be described phenomenologically by a two-dimensional shifted inverse Gaussian (IG) distribution \citep{Kopczynski/etal/2012a}.
The one-dimensional shifted IG is defined by the probability density
\begin{align}
   & g(x \given \mu, \lambda, o)\nonumber \\
   & \coloneq \begin{cases}
     0 & \text{if } x \leq o, \\
     \sqrt{\lambda / (2 \pi (x - o)^3)} \cdot
     \exp \left(-\lambda \frac{(x - o - \mu)^2}{2 \mu^2 (x - o)}
     \right)  & \text{otherwise,}\label{eq:inversegauss}
     \end{cases}
\end{align}
where~$o$ is an offset value.
The density of a peak is
\begin{equation}
p(r, t \given \theta) 
= v \cdot g(t \given \mu_{\text{T}}, \lambda_{\text{T}}, o_{\text{T}}) \cdot g(r \given \mu_{\text{R}}, \lambda_{\text{R}}, o_{\text{R}}),
\end{equation}
where $v$ is the volume of the peak and $\theta = (\mu_{\text{T}}, \lambda_{\text{T}}, o_{\text{T}}, \mu_{\text{R}}, \lambda_{\text{R}}, o_{\text{R}}, v)$.

Since the parameters~$\mu,\lambda, o$ vary strongly on similar shapes, it is more intuitive to describe the function in terms of three descriptors, the mean~$\mu'$, the standard deviation~$\sigma$ and the mode~$m$.
There is a bijection between~$(\mu,\lambda, o)$ and~$(\mu', \sigma, m)$ given by
\begin{align*}
  \mu'   &= \mu + o, \\
  \sigma &= \sqrt{\mu^3 / \lambda},\\
  m      &= \mu \left(\sqrt{ 1 + (9\mu^2)/(4\lambda^2)} - (3\mu)/(2\lambda) \right) + o,
\end{align*}
and the model parameters~$(\mu,\lambda, o)$ can be uniquely recovered from these descriptors.

The descriptors are drawn uniformly from the following intervals (the unit for retention times is~s, the unit for RIMs is~$\text{Vs}/\text{cm}^2$, and volumes~$v$ are given in arbitrary volume units):
\begin{center}\begin{tabular}{r@{ }c@{ }l}
$m_{\text{T}}$ & $\in$ & $[0.551, 1.015]$\\
$\sigma_{\text{T}}$ & $\in$ & $[0.0046, 0.00174]$\\
$\mu'_{\text{T}}$ & $\in$ & $[m_{\text{T}} + 0.00058, m_{\text{T}} + 0.0029]$\\
$m_{\text{R}}$ & $\in$ & $[25, 250]$ \\
$\sigma_{\text{R}}$ &$\in$& $[4, 7.5]$ \\
$\mu'_{\text{R}}$ & $\in$ & $[m_{\text{R}} + 0.5,  m_{\text{R}} + 2.5]$\\
$v$ & $\in$ & $[1.45, 14.5]$\\
\end{tabular}\end{center}


A simulated IMSC is generated as follows.
A number~$C$ of peaks is determined randomly from an interval (e.g., 5--10).
Peak descriptors are randomly drawn for each peak from the above intervals, and the model parameters~$\theta_j$ are computed for~$j=1,\dots,C$.
The IMSC~$M$ is generated by setting $M_{r,t} := \sum_{j=1}^C\, p(r, t \given \theta_j)$ for $r \leq |R|, t \leq |T|$.

\paragraph{Generating noisy IMSCs}
Starting with a peak-containing IMSC, we add normally distributed noise with parameters $\mu_{\text{N}} = 0.8$, $\sigma_{\text{N}} = 2.0$ (both in signal units), estimated from background noise of original IMSCs, to each data point~$M_{r,t}$.

Additionally, due to the device properties, the intensities in a spectrum are oscillating with a low frequency $f_r \in [1000, 6000]\,\text{Hz}$ that may change with retention time~$r$.
Thus we add $i \cdot \sin(\frac{U}{f_r \cdot l^2} \cdot T_t)$ to~$M_{r,t}$ where~$i$ is the intensity factor (note: $\frac{4380}{12^2}$ is our factor to compute drift times in RIMs).
Our tests showed that~$i \approx 1$ in practice.
The IMSC with added noise is called~$M'=(M'_{r,t})$.

\paragraph{Comparing IMSCs}
As a similarity measure betweens IMSCs~$M$ and~$N$, we use cosine similarity,
\[
\mathcal{S}(M, N) \coloneq \frac{\sum_{r,t}\, M_{r,t} \cdot N_{r,t}}{\sqrt{\sum_{r,t}\, M_{r,t}^2} \cdot \sqrt{\sum_{r,t}\, N_{r,t}^2}}.
\]
We have~$\mathcal{S} \in [-1,1]$, where~$\mathcal{S} = 1$ means both that matrices are identical,~$\mathcal{S} = 0$ means that the values are ``orthogonal'' and~$\mathcal{S} = -1$ means that~$M_{x, y} = -N_{x, y}$.
In fact, the similarity measure is the cosine of the angle between the IMSCs when interpreted as vectors.

\subsection{Denoising}
\label{sec:eval:denoising}

\begin{figure}[t]\centering
\includegraphics[width=0.98\columnwidth]{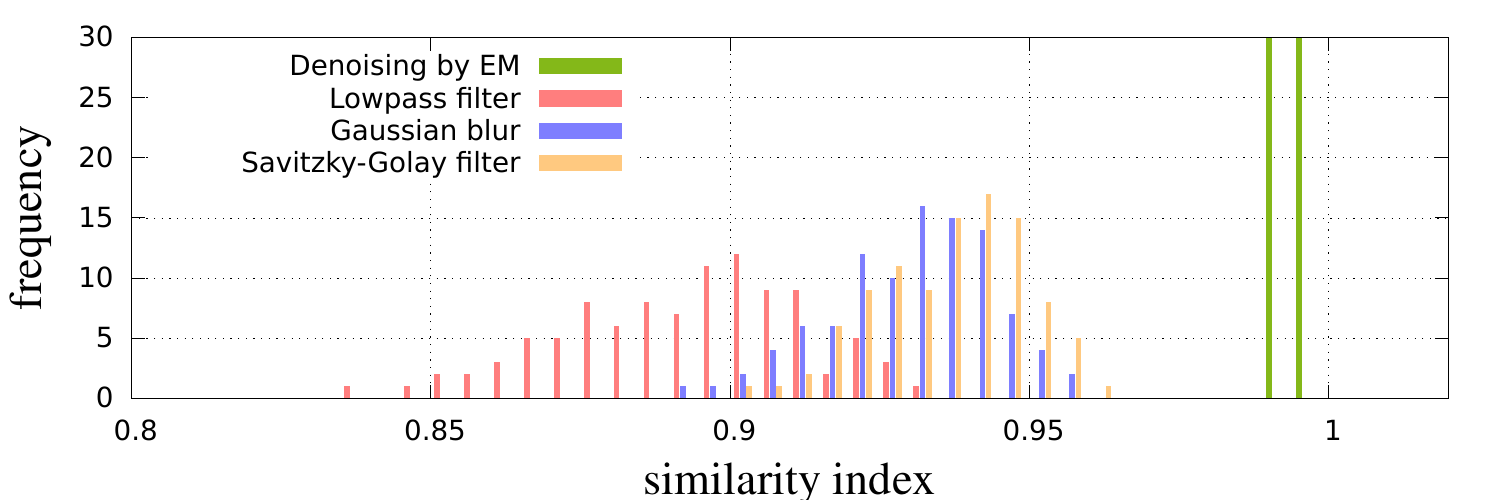}
\caption{Histogram of different methods' cosine similarity scores by comparing 100 original IMSCs with their denoised counterparts after adding simulated noise.}
\label{fig:denoisingres}
\end{figure}

We compared our method with current denoising and smoothing methods:
(1) Gaussian smoothing, (2) a Savitzky-Golay filter and (3) a low-pass filter utilizing the fast Fourier transform.

We first set up 100 different simulated IMSCs of~$800$ retention time points and~$2500$ RIM points, with 5--10 peaks, where the number of peaks is chosen randomly in this range, as described in Section~\ref{sec:eval:general}. 
These IMSCs are called~$M_i$, $i=1,\dots,100$.
We then add normal and sinusoid noise to each IMSC to obtain~$M'_i$, $i=1,\dots,100$.
We denoise the~$M'_i$ using our algorithm and the three above methods.
Let the resulting matrices be~$M^+_{i,\text{E}}$ (our EM~algorithm),~$M^+_{i,\text{L}}$ (low-pass filter),~$M^+_{i,\text{G}}$ (Gaussian smoothing) and~$M^+_{i,\text{S}}$ (Savitzky-Golay).
We compare each of these resulting matrices to the initial, noise-free matrix~$M_i$ using the cosine similarity measure described above.

We compute the cosine similarity~$\mathcal{S}(M_i,M^+_{i,A})$ between original and denoised IMSC.
with each algorithm~$A\in\{\text{E,G,L,S}\}$.
We show the histograms of the cosine similarity score of these 100 test cases in Figure~\ref{fig:denoisingres}.
The noisy IMSCs denoised with EM~denoising achieve higher similarity scores than by the other methods.

\subsection{Baseline Correction}
\label{sec:eval:baseline}

\begin{figure}[t]\centering
\includegraphics[width=0.98\columnwidth]{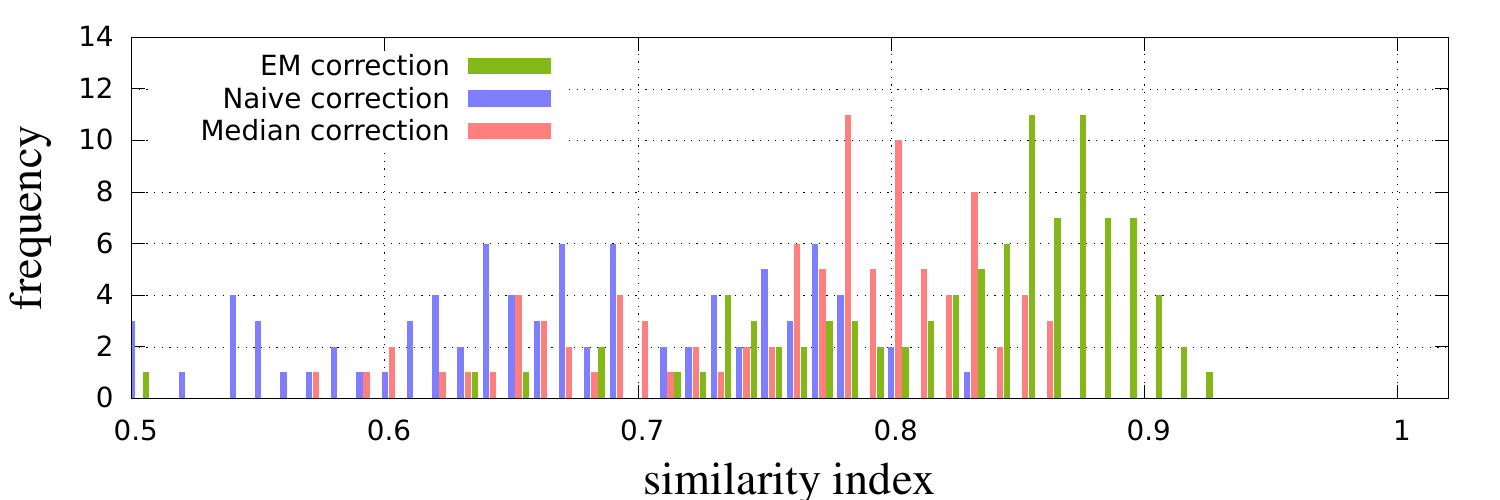}
\caption{Histogram of cosine similarity between initial simulated IMSCs and after baseline correction of the baseline-modified IMSCs with different algorithms.}
\label{fig:baselineresults}
\end{figure}

We compare the EM~baseline correction from Section~\ref{sec:baseline} with two additional methods.
\begin{enumerate}
\item
The first method (``naive'') subtracts a spectrum containing only baseline points from all remaining spectra.
Typically the first spectrum in an IMSC (captured after~$100\,\text{ms}$) consists only of a baseline, since even the smallest analytes need some time to pass the MCC.
After subtraction all negative values are set to zero.
\item
The second method (``median'') computes the median in every chromatogram separately and subtracts it from all values in the chromatogram.
Resulting negative values are set to~$0$.
\end{enumerate}

We simulate IMSCs~$M_i$ with 5--10 peaks for~$i=1,\dots,100$, and add normally distributed noise with an overlayed sinusoidal wave, as described in Section~\ref{sec:eval:general}, and then add a baseline to each spectrum in the IMSC based on the following considerations.
\begin{enumerate}
\item In theory, the amount of molecules getting ionized before entering the drift tube (and hence the sum over all intensities within a spectrum) is constant over spectra.
In practice, the amount varies and is observed to be normally distributed with a mean of about~$60\,000$ signal units and a standard deviation of about~$600$ signal units.
The signal intensity sum~$\tau_{r}$ for the~$r$-th spectrum is obtained by drawing from this normal distribution.
\item To obtain the signal intensity for non-peaks, we subtract the signal intensity consumed by simulated peaks in this spectrum. 
Let~$j$ index peaks in a given IMSC, and let~$p_{j}(r,t)$ be the signal intensity of the~$j$-th peak at coordinates~$(r,t)$.
Thus we compute $\tau'_r := \tau_r - \sum_{t}\, \sum_{j}\, p_j(r,t)$.
We repeat this process to obtain~$\tau'_{i,r}$ for every IMSC indexed by~$i$.
\item The baseline~$B_i(t)$ is modeled by two Inverse Gaussian distributions, one for the RIP~($\alpha$ component) and one for the heavy tail~($\beta$ component) of the RIP (cf.\ the work by \cite{bader2008preprocessing}, who used the log-normal distribution for the tail).
\[
B(t) := \omega \cdot g(t \given \mu_{\alpha}, \lambda_{\alpha}, o_{\alpha}) + (1 - \omega) \cdot g(t \given \mu_{\beta}, \lambda_{\beta}, o_{\beta}),
\]
where~$g$ was defined in Eq.~\eqref{eq:inversegauss} and the parameters are set to or uniformly drawn from
\begin{center}\begin{tabular}{r@{ }c@{ }l}
$\mu_{\alpha}$ & $=$ & $0.174$\\
$\lambda_{\alpha}$ & $\in$ & $[0.087, 0.127]$\\
$o_{\alpha}$ & $=$ & $0.443$\\
$\mu_{\beta}$ & $=$ & $0.127$ \\
$\lambda_{\beta}$ & $\in$ & $[23.2, 29]$\\
$o_{\beta}$ & $=$ & $0.353$\\
$\omega$ & $\in$ & $[0.6, 0.7]$\\
\end{tabular}\end{center}
where all units are~$\text{Vs}/\text{cm}^2$, except for~$\omega$. We repeat this process for every IMSC to obtain~$B_i(t)$ for~$i=1,\dots,100$.
\item The IMSC with baseline~$M'_i$ is obtained from the original IMSC~$M_i$ by
\[
M'_{i,r,t} = M_{i,r,t} + \tau_{i,r}' \cdot B_i(t)
\]
for $i=1,\dots,100$, $r\in R$, $t\in T$.
\end{enumerate}
We apply the three algorithms (EM, naive, median) to obtain~$M^+_{i,\text{E}}$ (for our EM-based method),~$M^+_{i,\text{N}}$ (naive) and~$M^+_{i,\text{M}}$ (median) and measure the cosine similarity~$\mathcal{S}(M_i, M^+_{i,A})$ for all~$i=1,\dots,100$ and algorithms~$A\in\{E,M,N\}$ and plot the results in Figure~\ref{fig:baselineresults}.
In average, the EM~baseline correction performs best in terms of cosine similarity.
Note that there is no explicit denoising performed in this experiment.

\subsection{Clustering}
\label{sec:eval:clustering}

\begin{figure}[t]\centering
\includegraphics[width=0.48\columnwidth]{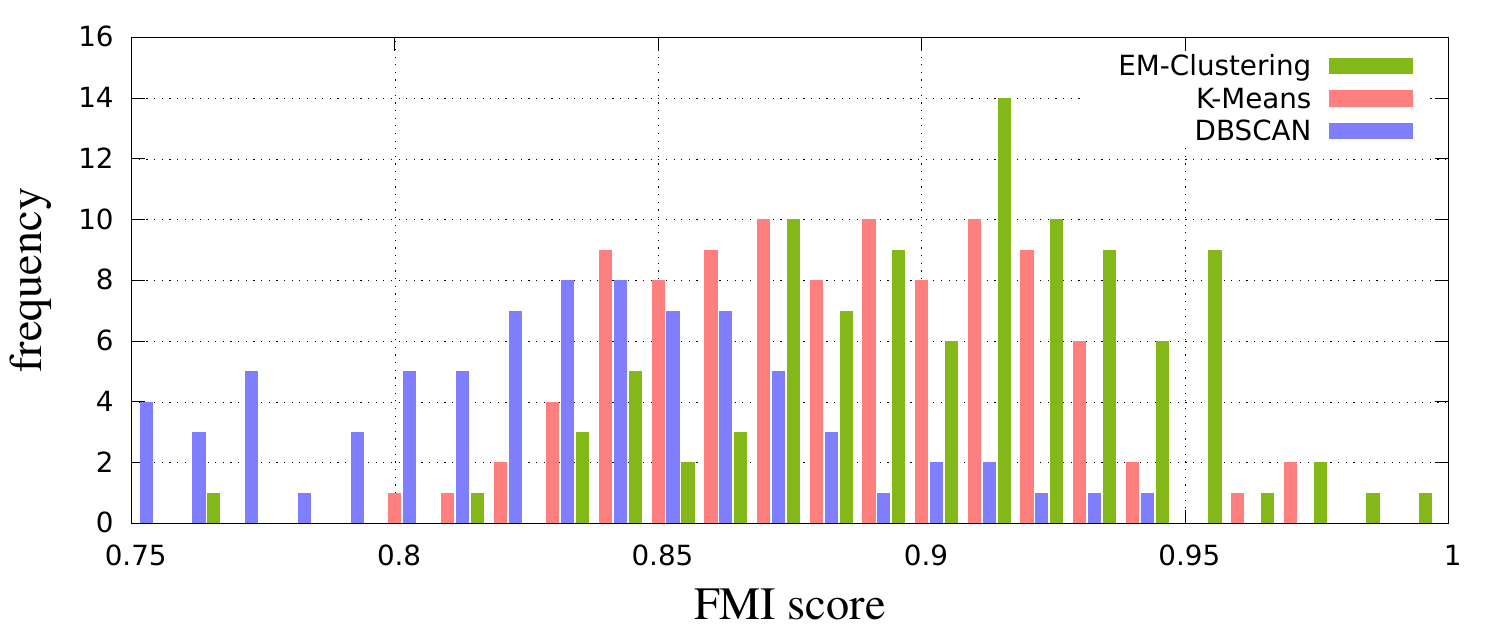}
\includegraphics[width=0.48\columnwidth]{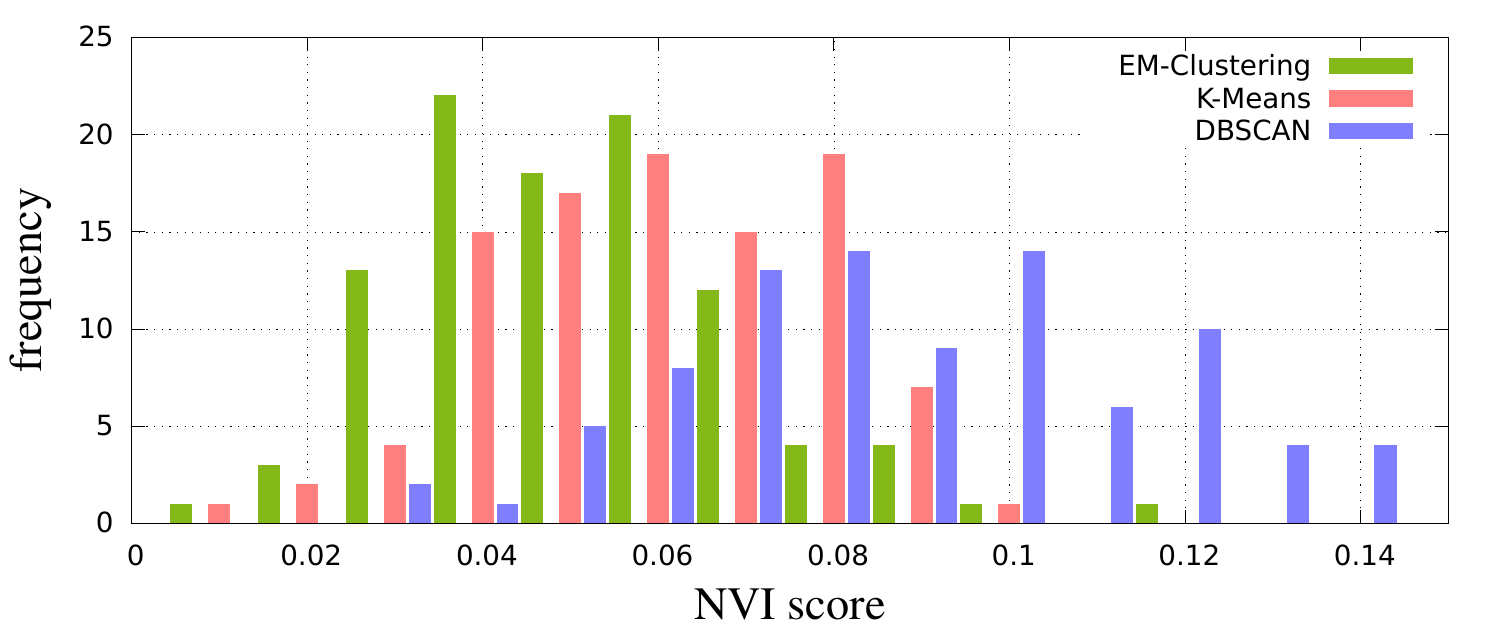}
\caption{
Histograms of Fowlkes-Mallows index (FMI; higher is better) and normalized variation of information (NVI; lower is better) comparing 100 simulated measurements containing partitioned peak locations with their clusters produced by the different methods.
EM~clustering achieves slightly better results than DBSCAN.
$K$-means can also come up.
}
\label{fig:clusterresultswn}
\end{figure}

\begin{figure}[t]\centering
\includegraphics[width=0.48\columnwidth]{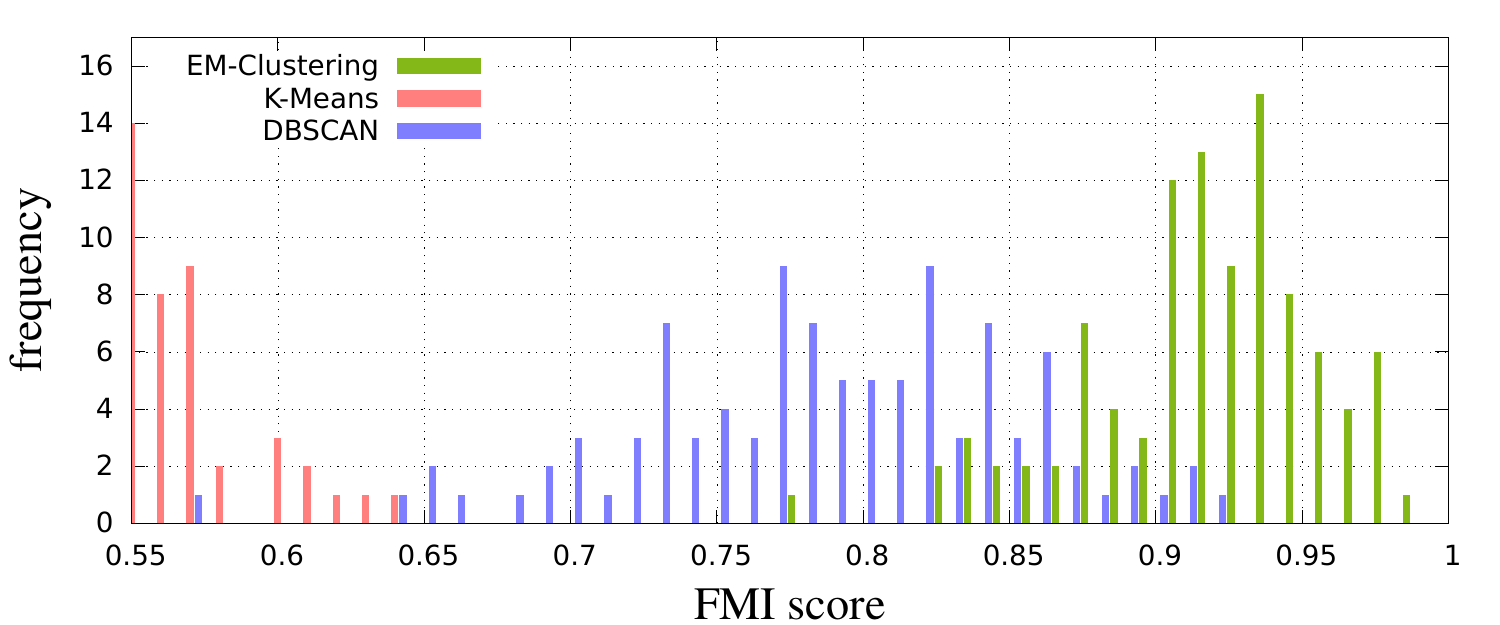}
\includegraphics[width=0.48\columnwidth]{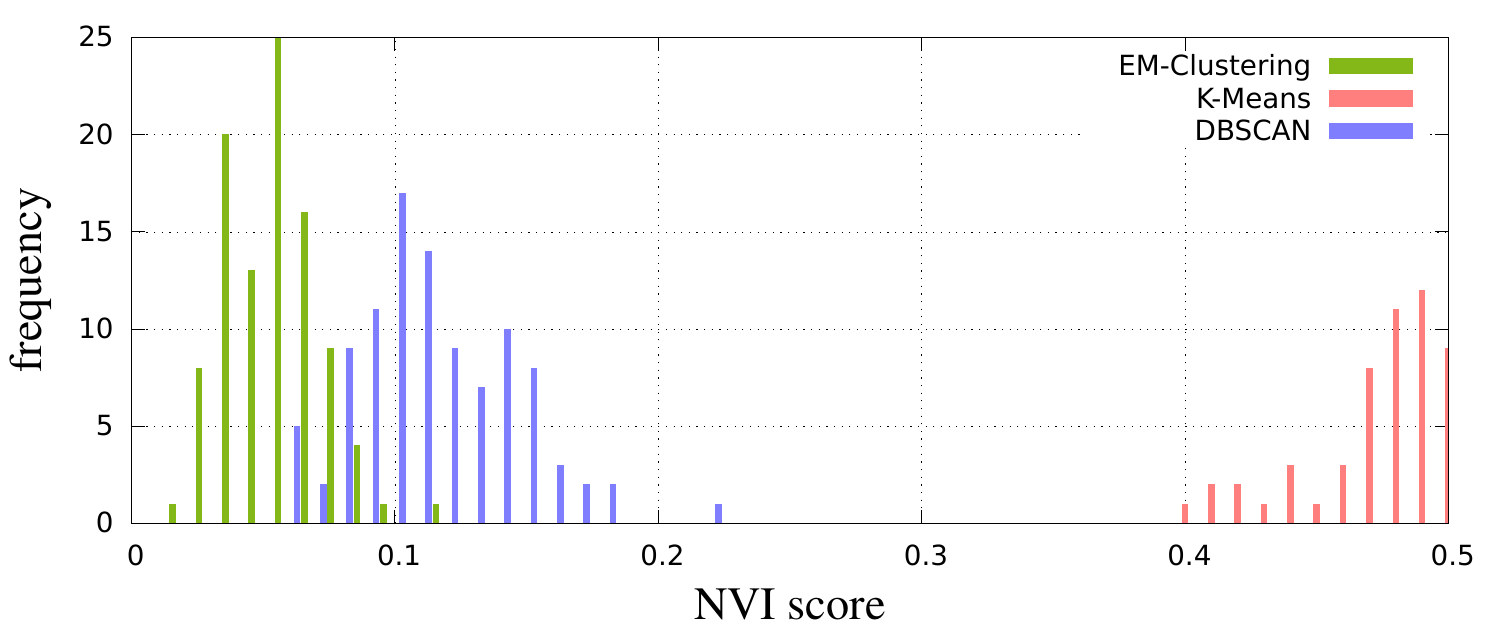}
\caption{
Histograms similar to Figure~\ref{fig:clusterresultswn}, but in a more realistic scenario with noise added.
Again, EM~clustering achieves in both score better results in average.
$K$-means performs less well because of insufficient choice of starting points when having much noise.
}
\label{fig:clusterresults}
\end{figure}

To evaluate peak clustering methods, we simulate peak locations according to locations in real MCC/IMS datasets, together with the true partition~$\mathcal{P}$ of peaks.

Most of the detected peaks appear in a small dense area early in the measurement, since many volatile compounds have a small chemical structure like ethanol or acetone.
Remaining peaks are distributed widely, which is referred to as the sparse area.
The areas have the following boundaries(in units of~($\text{Vs}/\text{cm}^2$, s) from lower left to upper right point, cf.\ Figure~\ref{fig:visualized-imsc}:
\begin{center}\begin{tabular}{r@{ }l}
measurement: & $(0, 0), (1.45, 600)$\\
dense area: & $(0.5, 4), (0.7, 60)$\\
sparse area: & $(0.5, 4), (1.2, 450)$
\end{tabular}\end{center}

Peak clusters are ellipsoidal and dense.
From \cite{bodeker2008peak} we know the minimum required distance between two peaks in order to be identified as two separate compounds.
We simulate peak cluster centroids, 30 in the dense area and 20 in the sparse area, all picked randomly and uniformly distributed.
We then randomly pick the number of peaks per cluster.
We also randomly pick the distribution of peaks within a cluster.
Since we do not know the actual distribution model, we decided to simulate with three models: normal~(n), exponential~(e) and uniform~(u) distribution with the following densities:
\begin{align*}
 & f_{\text{n}}(r, t \given \mu_\text{t}, \sigma_\text{t}, \mu_\text{r}, \sigma_\text{r}) \\
 &\quad = \mathcal{N}(t \given \mu_\text{t}, \sigma_\text{t}) \cdot \mathcal{N}(r \given \mu_\text{r}, \sigma_\text{r}) \\
 & f_{\text{e}}(r, t \given \mu_\text{t}, \lambda_\text{t}, \mu_\text{r}, \lambda_\text{r}) \\
 &\quad = \lambda_\text{t} \lambda_\text{r} \exp \big(- (\lambda_\text{t} |t-\mu_\text{t}| + \lambda_r |r-\mu_\text{r}|) \big) / 4 \\
 & f_{\text{u}}(r, t \given  \mu_\text{t}, \rho_\text{t},  \mu_\text{r}, \rho_\text{r}) \\
 &\quad = \begin{cases}
     (\pi \rho_\text{t} \rho_\text{r} )^{-1} & \text{if } \frac{|t-\mu_\text{t}|^2}{\rho_\text{t}^2} + \frac{|r-\mu_\text{r}|^2}{\rho_\text{r}^2} \leq 1 \\
     0 & \text{otherwise}
     \end{cases}
\end{align*}
Here $(\mu_\text{t}, \mu_\text{r})$ is the coordinate of the centroid with RIM in $\text{Vs}/\text{cm}^2$ and retention time in~s.
For the normal distribution, $\sigma_{\text{t}}=0.002$ and $\sigma_{\text{r}}=\mu_\text{r} \cdot 0.002 + 0.2$.
For exponential distribution, $\lambda_{\text{t}} = (1.45 \cdot 2500)^{-1}$ (reduced mobility width for in single cell within~$M$) and~$\lambda_{\text{r}} = (\mu_\text{r} \cdot 0.002 + 0.2)^{-1}$.
For the uniform distribution, we use an ellipsoid with radii $\rho_{\text{t}} = 0.006$ and $\rho_{\text{r}} = \mu_\text{r} \cdot 0.02 + 1$.

We compared the EM~clustering with two common clustering methods, namely~$K$-means and DBSCAN.
Since~$K$-means needs a fixed~$K$ for the number of clusters and appropriate start values for the centroids, we decided to take~$K$-means++ (described by~\cite{arthur/etal/2007}) for estimating good starting values and give it an advantage by assigning the true number of partitions.
DBSCAN has the advantage that it does not need a fixed number of clusters, but on the other hand it has some disadvantages.
It finds clusters with non-linearly separable connections, but we assume that the partitions obey a kind of model with convex hull.
On the other hand it yields no parameters describing the clusters.
Such parameters can be very important when using the clusters as features for a consecutive classification.

To measure the quality of the clustering~$\mathcal{C}$ we take two measures in consideration: the Fowlkes-Mallows index (FMI) first described by~\cite{fowkles/etal/1983} as well as the normalized variation of information (NVI) score introduced by~\cite{reichart/etal/2009}.

For the FMI one has to consider all pairs of data points.
If two data points belong into the same true partition of~$\mathcal{P}$, they are called~\emph{connected}.
Accordingly, a pair of data points is called~\emph{clustered} if they are clustered together by the clustering method we want to evaluate.
Pairs of data points, which are marked as connected as well as clustered, are referred to as true positives (TP).
False positives (FP, not connected but clustered) and false negatives (FN, connected but not clustered) are computed, analogously.
The FMI is the geometric mean of precision and recall, let \hbox{$\text{FMI}(\mathcal{P}, \mathcal{C}) \coloneq \sqrt{TP / (TP + FP) \cdot TP / (TP + FN)}$} where~$\mathcal{P}$ is the partition set and~$\mathcal{C}$ the clustering.
Since \hbox{$\text{FMI}(\mathcal{P}, \mathcal{C}) \in [0, 1]$}, $\text{FMI}(\mathcal{P}, \mathcal{C}) = 0$ means no similarity between both clusterings and $\text{FMI}(\mathcal{P}, \mathcal{C}) = 1$ means that the clusterings agree completely.
Although the FMI determines the similarity between two clusterings, it yields unreliable results when the number of clusters in both clusterings differs significantly.

Thus we use a second measure that considers clusters sizes only, the normalized variation of information (NVI).
To compute the NVI, an auxiliary~$|\mathcal{P}| \times |\mathcal{C}|$-dimensional matrix~$A = (a_{i,j})$ has to be set up.
Thereby~$a_{i,j}$ determines the number of data points within partition~$i$ that are assigned to cluster~$j$.
Using entropies, we can now determine the NVI score.
Define
\begin{align*}
H(\mathcal{P}) &\coloneq -\sum_{i \leq |\mathcal{P}|} \frac{ \sum_{j \leq |\mathcal{C}|} a_{i,j} }{n} \log \frac{ \sum_{j \leq |\mathcal{C}|} a_{i,j} }{n}r, \\
H(\mathcal{C}) &\coloneq -\sum_{j \leq |\mathcal{C}|} \frac{ \sum_{i \leq |\mathcal{P}|} a_{i,j} }{n} \log \frac{ \sum_{i \leq |\mathcal{P}|} a_{i,j} }{n}, \\
H(\mathcal{P} | \mathcal{C}) &\coloneq -\sum_{j \leq |\mathcal{C}|} \sum_{i \leq |\mathcal{P}|} \frac{a_{i,j}}{n} \log \frac{a_{i,j}}{\sum_{i' \leq |\mathcal{P}|}a_{i',j}},  \\
H(\mathcal{C} | \mathcal{P}) &\coloneq -\sum_{j \leq |\mathcal{C}|} \sum_{i \leq |\mathcal{P}|} \frac{a_{i,j}}{n} \log \frac{a_{i,j}}{\sum_{j' \leq |\mathcal{C}|}a_{i,j'}},  \\
NVI(\mathcal{P}, \mathcal{C}) &\coloneq  
\begin{cases}
    \frac{H(\mathcal{P} | \mathcal{C}) + H(\mathcal{C} | \mathcal{P})}{H(\mathcal{P})} & \text{if } H(\mathcal{P}) \neq 0, \\
    H(\mathcal{C})              & \text{otherwise}
\end{cases}
\end{align*}
where~$n$ is the number of data points.
$NVI(\mathcal{P}, \mathcal{C}) = 0$ means no variation between original partition and clustered data.
An FMI score~$= 1$ and NVI score~$ = 0$ indicates a perfect clustering.

For the first test we generated 100 sets of data points where the partitions is known, as previously described.
In the second step performed an EM~clustering as well as~$K$-means and DBSCAN for every set.
Finally we computed the both scores FMI and NVI for all sets.
Our results show that even with the unfair~$K$-means our EM~clustering performs best in terms FMI and NVI score.
It achieves in average best results, Figure~\ref{fig:clusterresultswn} shows two histograms of both FMI and NVI for all three methods.
Since this scenario is little realistic, we performed a second test.
The difference to the first test is that we insert 200 equally distributed peaks randomly into the measurement area.
All these peaks are singletons within the partition set.
We denote the additional peaks as noise.
After performing the second test, we can see that EM~clustering still achieves best results in average, whereas  $K$-means completely fails although we forward the correct~$K$, because of insufficient determination of start points and no noise handling.
All FMI and NVI scores from the second test are plotted as a histogram in Figure~\ref{fig:clusterresults}.
%


\section{Discussion and Conclusion}
\label{sec:discussion}

We have presented three novel methods for certain problems i.e. denoising, baseline correction and clustering.
All methods utilize a modified version of the EM~algorithm for a deconvolution of mixture models.
Since our research is located in spectra analysis of ion MCC/IMS devices, these methods are adjusted for this application field, but can easily be adapted for other purposes.
In all tests our methods performed with best results.
Because of lack of the truth behind original measurements, we simulated test data using properties of real MCC/IMS measurements.
All methods are being applied for automated breath gas analysis to improve the accuracy of disease prediction, as previously evaluated by~\cite{hauschild2013eval}.

Supplementary material (parameter lists for all denoising and baseline correction tests as well as peak lists for clustering) are available at \url{http://www.rahmannlab.de/research/ims}.


\paragraph*{Acknowledgements}
DK, SR are supported by the Collaborative Research Center (Sonderforschungsbereich,~SFB) 876 ``Providing Information by Resource-Constrained Data Analysis'' within project TB1, see \url{http://sfb876.tu-dortmund.de}.
\bibliographystyle{natbib}

\end{document}